\documentclass[11pt]{article}

\usepackage[margin=1in]{geometry}
\usepackage{setspace}

\usepackage{graphicx}
\usepackage{multirow}
\usepackage{amsmath,amssymb,amsfonts}
\usepackage{amsthm}
\usepackage{mathrsfs}
\usepackage[title]{appendix}
\usepackage{xcolor}
\usepackage{textcomp}
\usepackage{booktabs}
\usepackage{algorithm}
\usepackage{algorithmicx}
\usepackage{algpseudocode}
\usepackage{listings}
\usepackage{subcaption}
\usepackage[authoryear,round]{natbib}  % natbib在这
\usepackage{hyperref}  % hyperref最后

\theoremstyle{plain}

\theoremstyle{definition}

\theoremstyle{remark}

\begin{document}

\title{Mapping Drivers of Greenness: Spatial Variable Selection for MODIS Vegetation Indices}

\author{
Qishi Zhan\thanks{Department of Mathematical and Statistical Sciences, Marquette University} \and
Cheng-Han Yu\thanks{Department of Mathematical and Statistical Sciences, Marquette University} \and
Yuchi Chen\thanks{Department of Computer Science, Georgia Institute of Technology} \and
Zhikang Dong\thanks{Department of Applied Mathematics and Statistics, Stony Brook University} \and
Rajarshi Guhaniyogi\thanks{Department of Statistics, Texas A\&M University}
}

\maketitle

\begin{abstract} Understanding how environmental drivers relate to vegetation condition motivates spatially varying regression models, but estimating a separate coefficient surface for every predictor can yield noisy patterns and poor interpretability when many predictors are irrelevant. Motivated by MODIS vegetation index studies, we examine predictors from spectral bands, productivity and energy fluxes, observation geometry, and land surface characteristics. Because these relationships vary with canopy structure, climate, land use, and measurement conditions, methods should both model spatially varying effects and identify where predictors matter. We propose a spatially varying coefficient model where each coefficient surface uses a tensor product B-spline basis and a Bayesian group lasso prior on the basis coefficients. This prior induces predictor level shrinkage, pushing negligible effects toward zero while preserving spatial structure. Posterior inference uses Markov chain Monte Carlo and provides uncertainty quantification for each effect surface. We summarize retained effects with spatial significance maps that mark locations where the 95 percent posterior credible interval excludes zero, and we define a spatial coverage probability as the proportion of locations where the credible interval excludes zero. Simulations recover sparsity and achieve prediction. A MODIS application yields a parsimonious subset of predictors whose effect maps clarify dominant controls across landscapes.
\end{abstract}


\section{Introduction}\label{sec:intro}
The western United States is undergoing rapid and unprecedented ecological change, with vegetation increasingly exposed to severe drought stress and heightened wildfire risk \citep{abatzoglou2016impact,williams2019observed,williams2020large}. These changes carry enormous economic costs: wildfires in the region are estimated to cost the U.S. economy \$394–\$893 billion annually, with damages exceeding \$90 billion during 2017–2021 alone. Accurately identifying which environmental factors govern vegetation health and productivity is therefore crucial for predicting ecosystem responses and designing effective adaptation strategies \citep{running2004continuous, pettorelli2005using}.

This challenge is amplified by the region’s extreme environmental heterogeneity. The western United States spans from coastal rainforests receiving over 3000 mm of annual precipitation to interior deserts with less than 200 mm \citep{diffenbaugh2015anthropogenic}, leading to fundamentally different ecological constraints. At high elevations, temperature often limits productivity, whereas in arid lowlands, moisture availability is the dominant control \citep{sun2024alternating,mukhtar2025elevation}. Effective vegetation monitoring in such a setting requires identifying which environmental variables exert substantively nonzero effects on vegetation across space, and determining where within the domain those effects are credibly nonzero.

We address this problem using a predictive modeling framework for MODIS Enhanced Vegetation Index (EVI) data from the western United States, leveraging a rich suite of environmental predictors: spectral reflectance bands (red, near-infrared, blue, mid-infrared), ecosystem productivity and energy fluxes (Gross Primary Productivity, Latent Heat Flux), observation geometry (view zenith angle, sun zenith angle, relative azimuth angle), and land surface characteristics (land cover type). Standard linear regression models assume that coefficients are spatially constant, potentially obscuring important geographic variation in these relationships \citep{fotheringham2002geographically}. At the same time, not all candidate drivers have effects that are credibly different from zero across the region. Systematically screening predictors for non-negligible effects improves interpretability and directs scientific focus toward drivers that truly matter. Predictors with spatially extensive, credibly nonzero effects motivate spatially targeted interpretation and monitoring efforts, whereas predictors with negligible influence can often be excluded or modeled more simply, reducing monitoring and modeling complexity without sacrificing predictive accuracy.

Spatially varying coefficient (SVC) models are well suited to this problem because they allow regression relationships to change smoothly across space, producing coefficient surfaces that capture geographic variation in environmental effects \citep{brunsdon1996geographically,fotheringham2017multiscale,comber2024multiscale}. These approaches have illuminated how environmental relationships vary across elevation gradients, climate zones, and land-use types, thereby supporting targeted interventions where specific drivers exert the greatest influence \citep{gelfand2003spatial,finley2011comparing}.
Among popular SVC methods, geographically weighted regression (GWR) \citep{brunsdon1996geographically} computes local coefficients via distance-weighted kernels; multiscale GWR (MGWR) \citep{fotheringham2017multiscale} extends this by allowing each predictor to operate at its own spatial scale through variable-specific bandwidths; and the geographical Gaussian process generalized additive model (GGP-GAM) \citep{comber2024multiscale} embeds spatial smoothers within a generalized additive model framework, yielding a unified model with strong theoretical support. These methods are highly effective at capturing spatial variation when it is present. However, identifying which predictors truly have practically important spatial effects, and which have negligible effects that can be screened out, remains a distinct challenge that requires integrated spatial variable selection \citep{wheeler2005multicollinearity,finley2011comparing}. While there is a well-developed literature on spatial variable selection in imaging \citep{goldsmith2014smooth, smith2003assessing}, in econometrics \citep{scheel2013bayesian}, in public health \citep{choi2018bayesian} and in environmental studies \citep{meyer2019importance}, to the best of our knowledge, no prior study provides a principled spatial variable selection framework specifically aimed at identifying spatially varying environmental drivers of vegetation indices.

\color{black}{
To address this gap, we develop a Bayesian spatially varying coefficient (SVC) regression model for EVI that accommodates a broad set of environmental drivers and enables location-specific identification of important versus unimportant predictors. Each predictor-specific coefficient surface is expressed using a tensor product B-spline basis, and the associated basis coefficient vector is endowed with a Bayesian group lasso prior \citep{yuan2006model, xu2015bayesian}. This prior induces predictor-level shrinkage: predictors with negligible influence are encouraged to have coefficient surfaces pulled toward zero, while those with practically important effects retain nontrivial spatial patterns. Posterior inference is carried out via Markov chain Monte Carlo, yielding full posterior distributions and pointwise 95\% credible intervals for each coefficient surface. For each predictor, we then construct a spatial significance map by determining, at every observed location, whether the 95\% credible interval includes zero. This provides a spatially explicit characterization of predictor importance for EVI over the entire domain. To summarize importance at the regional scale, we introduce a spatial coverage probability (SCP) for each predictor, defined as the proportion of locations where its coefficient is credibly nonzero. Together, the spatial significance maps and SCP summaries yield spatially-varying significance of each predictor to predict EVI in our MODIS-based vegetation analysis of the western United States.
\color{black}

\color{black} Application to MODIS vegetation data demonstrates the practical value of our framework. We find that near-infrared reflectance, red reflectance, and gross primary productivity exhibit spatially extensive, credibly nonzero effects across diverse environmental gradients, marking them as primary, large-scale drivers of vegetation condition. In contrast, mid-infrared reflectance and land cover type have coefficients that are not credibly different from zero over most of the domain, indicating that their added explanatory power for EVI is limited in this setting. These findings are consistent with established vegetation science, where NIR and red bands form the basis of most greenness indices and are tightly linked to photosynthetic activity and productivity \citep[e.g.,][]{tucker1979red,huete2002overview,running2004continuous,pettorelli2005using}. Predictors with widespread, credibly nonzero effects warrant focused scientific interpretation and ongoing monitoring, whereas predictors with negligible influence can be omitted or modeled more parsimoniously without loss of interpretability or predictive ability.
\color{black} 

The rest of the paper is organized as follows. %Section~\ref{sec:background} reviews spatially varying coefficient models and establishes the need for integrated variable selection in ecological applications. 
Section~\ref{sec:methods} develops the Bayesian additive model framework and spatial significance mapping approach. Section~\ref{sec:simulation} presents simulation studies demonstrating performance across varying sample sizes and dimensionalities under both independent and correlated predictor scenarios. Section~\ref{sec:application} analyzes MODIS vegetation data from the western United States. Section~\ref{sec:discussion} summarizes the contribution and discusses methodological extensions. Open source implementation is available at \url{https://github.com/Qishi7/BSGL}.

\section{Methods}\label{sec:methods}

\subsection{Model Specification}\label{subsec:model}
Ecological relationships often exhibit spatial heterogeneity driven by environmental gradients, resource availability, and biophysical constraints \citep{turner2005landscape,pickett1995landscape}. Spatially varying coefficient (SVC) models provide a framework for investigating such heterogeneity by allowing regression coefficients to vary over space. The model takes the form 
\begin{equation}\label{eq:gwr}
y({\boldsymbol z}_i) = \beta_0({\boldsymbol z}_i) + \sum_{j=1}^{m} x_{j}({\boldsymbol z}_i)\beta_j({\boldsymbol z}_i) + \varepsilon_i
\end{equation}
where $y({\boldsymbol z}_i)$ and ${\boldsymbol x}({\boldsymbol z}_i)=(x_1({\boldsymbol z}_i),...,x_m({\boldsymbol z}_i))^T$ represent the response and predictor variables, ${\boldsymbol z}_i  = (u_i, v_i)$ represents spatial coordinates and $\beta_j({\boldsymbol z}_i)$ are local regression coefficients. The idiosyncratic errors are taken to be independent and identically distributed (i.i.d.), $\varepsilon_i \stackrel{\text{i.i.d.}}{\sim} N(0, \sigma^2)$. The SVC employs a separate regression for each spatial location, with these local modeling approaches capturing how the relationship between vegetation indices and environmental drivers varies across the landscape. A standard Bayesian implementation specifies Gaussian process (GP) priors for the coefficient surfaces $\beta_j({\boldsymbol z})$ \citep{guhaniyogi2022distributed,guhaniyogi2023distributed}. While GP-based SVC models are effective for describing smoothly varying effects, they are less suitable for our MODIS vegetation application, where a central goal is to distinguish environmental drivers with substantively nonzero effects from those with negligible influence that should be screened out.

To facilitate both spatial smoothing and variable selection, we instead represent each coefficient surface $\beta_j({\boldsymbol z})$ via a finite basis expansion,
\begin{equation}\label{eq:ggp_basis}
\beta_j({\boldsymbol z}) = \sum_{l=1}^{L} \alpha_{jl} \psi_l({\boldsymbol z}),
\end{equation}
where \(\{\psi_l(\boldsymbol{z})\}_{l=1}^{L}\) is a chosen set of spatial basis functions and \(\alpha_{jl}\) are the corresponding basis coefficients. A wide range of basis systems have been proposed for spatial regression, including B-splines, radial basis functions, and wavelets \citep[e.g.,][]{eilers1996flexible,wahba1990spline,kammann2003geoadditive,ruppert2003semiparametric}. For our empirical analysis, we use tensor-product B-spline bases to flexibly approximate spatial coefficient surfaces while enabling shrinkage and selection at the predictor level, described in the next section.

\subsection{Hierarchical Prior with Group lasso}\label{subsec:prior}

Following \eqref{eq:gwr}, our proposed model takes the form:
\begin{equation}
\mathbf{y} = \sum_{j=1}^{m} \mathbf{X}_j\mathbf{\Psi}\boldsymbol{\alpha}_j + \boldsymbol{\varepsilon},
\end{equation}
where $\mathbf{y} = (y({\boldsymbol z}_1), \ldots, y({\boldsymbol z}_n))^T$ is the response vector, $\mathbf{X}_j = \text{diag}(x_{j}({\boldsymbol z}_1), \ldots, x_{j}({\boldsymbol z}_n))$ is an $n \times n$ diagonal matrix containing the $j$-th covariate values, $\mathbf{\Psi}$ is an $n \times L$ matrix with entries $\Psi_{il} = \psi_l(z_i)$ representing spatial basis functions evaluated at locations ${\boldsymbol z}_i$, $\boldsymbol{\alpha}_j = (\alpha_{j1}, \ldots, \alpha_{jL})^T$ is the coefficient vector for the $j$-th covariate, and $\boldsymbol{\varepsilon} \sim N(\mathbf{0}, \sigma^2\mathbf{I})$ with $\mathbf{I}$ an $n \times n$ identity matrix.

We employ a Bayesian group lasso prior that induces sparsity at the group level, where each group consists of all $L$ basis coefficients for a single predictor. This enables joint shrinkage of $\boldsymbol{\alpha}_j = (\alpha_{j1}, \ldots, \alpha_{jL})^\top$, \color{black} allowing the model to screen predictors with negligible effects by shrinking entire coefficient surfaces toward zero, while estimating coefficient surfaces for predictors with nonnegligible effects. \color{black}

The hierarchical specification is:
\begin{align}
\mathbf{y} \mid \{\boldsymbol{\alpha}_j\}_{j=1}^{m}, \sigma^2 &\sim N\left(\sum_{j=1}^{m} \mathbf{X}_j\mathbf{\Psi}\boldsymbol{\alpha}_j, \sigma^2\mathbf{I}\right) \nonumber, \\
\boldsymbol{\alpha}_j \mid \tau_j^2, \sigma^2 &\sim N(\mathbf{0}, \sigma^2\tau_j^2\mathbf{I}_L), \quad j = 1, \ldots, m \nonumber, \\
\tau_j^2 \mid \lambda^2 &\sim \text{Ga}\left(\frac{L+1}{2}, \frac{\lambda^2}{2}\right), \quad j = 1, \ldots, m \nonumber, \\
\sigma^2 &\sim \text{IG}(a_{\sigma}, b_{\sigma}), \qquad \lambda^2 \sim \text{Ga}(a_\lambda, b_\lambda).
\end{align}

This hierarchical structure induces a group lasso penalty. Integrating out the group-specific variance parameters $\tau_j^2$ produces a prior on $\boldsymbol{\alpha}_j$ that penalizes the $\ell_2$ norm $\|\boldsymbol{\alpha}_j\|_2 = \sqrt{\sum_{l=1}^L \alpha_{jl}^2}$ \citep{yuan2006model}. This group penalty shrinks entire coefficient vectors toward zero simultaneously, unlike standard lasso penalties that shrink individual coefficients independently. When applied to spatial basis coefficients, the prior encourages setting entire spatial surfaces $\beta_j({\boldsymbol z})$ to zero rather than reducing magnitudes at individual locations.

The parameter $\tau_j^2$ controls shrinkage intensity for predictor $j$. When $\tau_j^2$ is large, \color{black} the prior variance allows the basis coefficients to take substantial values, retaining coefficient surfaces with nonnegligible magnitude when supported by the data. \color{black} When $\tau_j^2$ approaches zero, all coefficients are shrunk toward zero, producing coefficients near zero across the spatial domain. The global parameter $\lambda^2$ governs overall sparsity. Larger values of $\lambda^2$ encourage smaller $\tau_j^2$ values through the gamma prior, inducing more aggressive shrinkage and \color{black} favoring models where fewer predictors have substantively nonzero coefficient surfaces. \color{black}

This hierarchical structure differs from standard Bayesian group lasso applications \citep{xu2015bayesian} where groups correspond to sets of related predictors or predefined variable clusters. In our spatial context, each group comprises the $L$ basis function coefficients $\boldsymbol{\alpha}_j$ that jointly define the $j$th predictor's spatial surface. \color{black} This spatial grouping structure enables the prior to distinguish between predictors with substantively nonzero coefficient surfaces and predictors with negligible influence on the response. \color{black}

\subsection{Posterior Inference: Spatial Significance Map and Spatial Coverage Probability }\label{subsec:inference}

We perform posterior inference using the developed Markov chain Monte Carlo (MCMC). The hierarchical structure leads to closed-form full conditional posterior distributions, enabling efficient sampling. Detailed derivations of the full conditionals are provided in Section 1 of the supplementary material.

The resulting MCMC algorithm alternates between these conditional updates until convergence, see Algorithm~\ref{algo:BSGL} for details. We assess convergence using four independent chains with the Gelman-Rubin diagnostic $\hat{R}$ \citep{gelman1992inference}, where values below 1.1 indicate satisfactory convergence, along with visual inspection of trace plots for proper mixing.

\begin{algorithm}[!t]
\caption{MCMC Sampling Algorithm}
\label{algo:BSGL}
\begin{algorithmic}[1]
\State \textbf{Initialize:} Hyperparameters $a_{\sigma}, b_{\sigma}, a_{\lambda}, b_{\lambda}$; starting values $\{\boldsymbol{\alpha}_j^{(0)}\}_{j=1}^m$, $\{(\tau_j^2)^{(0)}\}_{j=1}^m$, $(\sigma^2)^{(0)}$, $(\lambda^2)^{(0)}$
\For{$t = 1, 2, \ldots,$ until convergence}
    \For{$j = 1, \ldots, m$}
        \State $\boldsymbol{r}_j \gets \boldsymbol{y} - \sum_{k \neq j} \boldsymbol{X}_k\boldsymbol{\Psi}\boldsymbol{\alpha}_k^{(t-1)}$
        \State $\boldsymbol{V}_j^{(t)} = (\sigma^2)^{(t-1)}\left(\boldsymbol{\Psi}^{\top}\boldsymbol{X}_j^{\top}\boldsymbol{X}_j\boldsymbol{\Psi} + \frac{1}{(\tau_j^2)^{(t-1)}}\boldsymbol{I}_L\right)^{-1}$
        \State $\boldsymbol{m}_j^{(t)} = \frac{1}{(\sigma^2)^{(t-1)}}\boldsymbol{V}_j^{(t)}\boldsymbol{\Psi}^{\top}\boldsymbol{X}_j^{\top}\boldsymbol{r}_j$
        \State Sample $\left( \boldsymbol{\alpha}_j^{(t)} \mid \cdots \right) \sim N(\boldsymbol{m}_j^{(t)}, \boldsymbol{V}_j^{(t)})$
    \EndFor
    \For{$j = 1, \ldots, m$}
        \State Sample $\left( \gamma_j^{(t)} \mid \cdots \right) \sim \text{GIG}\left(-\frac{1}{2}, \frac{(\lambda^2)^{(t-1)}}{2}, \frac{(\boldsymbol{\alpha}_j^{(t)})^{\top}\boldsymbol{\alpha}_j^{(t)}}{2(\sigma^2)^{(t-1)}}\right)$
        \State $(\tau_j^2)^{(t)} \gets 1/\gamma_j^{(t)}$
    \EndFor
    \State $S^{(t)} = \frac{1}{2}\left\|\boldsymbol{y} - \sum_{j=1}^{m} \boldsymbol{X}_j\boldsymbol{\Psi}\boldsymbol{\alpha}_j^{(t)}\right\|_2^2 + \frac{1}{2}\sum_{j=1}^{m}\frac{(\boldsymbol{\alpha}_j^{(t)})^{\top}\boldsymbol{\alpha}_j^{(t)}}{(\tau_j^2)^{(t)}}$
    \State Sample $\left( \sigma^2 \mid \cdots \right) \sim \text{IG}\left(a_{\sigma} + \frac{n+mL}{2}, b_{\sigma} + S^{(t)}\right)$
    \State Sample $\left( \lambda^2 \mid \cdots \right) \sim \text{Ga}\left(a_{\lambda} + \frac{m(L+1)}{2}, b_{\lambda} + \frac{1}{2}\sum_{j=1}^m (\tau_j^2)^{(t)}\right)$
\EndFor
\end{algorithmic}
\end{algorithm}

%\subsection{Spatial Coverage Probability}\label{subsec:pip}

After obtaining posterior samples for the model parameters, we generate MCMC samples of the spatial coefficient functions. To identify predictors with spatially extensive, credibly nonzero effects, we examine pointwise posterior credible intervals across the study area. Specifically, for each predictor \(j\), we construct \((1-\alpha)100\%\) pointwise credible intervals (CIs) for \(\beta_j({\boldsymbol s})\) on a dense grid of locations $\mathcal{S} = \{{\boldsymbol s}_1,\dots, {\boldsymbol s}_G\}$ covering the domain. At each grid point \({\boldsymbol s}_g\), we check whether the credible interval excludes zero; if it does, we regard the effect of predictor \(j\) as credibly different from zero at that location. Plotting these classifications over space yields a spatial significance map for each predictor.

We then summarize the overall spatial extent of a predictor’s influence using the spatial coverage probability (SCP),
\begin{equation}
\text{SCP}_j = \frac{1}{G}\sum_{g=1}^{G} \mathbb{I}\{0 \notin \text{CI}_{1-\alpha}(\beta_j({\boldsymbol s}_g))\},
\end{equation}
where \(\mathbb{I}(\cdot)\) equals 1 if the credible interval at \({\boldsymbol s}_g\) excludes zero. Thus, \(\text{SCP}_j\) measures the proportion of the study region where predictor \(j\) has a credibly nonzero effect. Values close to 1 indicate that the predictor is important across most of the domain, while values near 0 suggest little or no influence. Notably, \(\text{SCP}_j\) is dependent on $\alpha$. Following the popular approach, we present results for $95\%$ CI for all simulations and vegetation data analysis. Section 4 of the supplementary file shows sensitivity to these choices with perturbing $\alpha$ moderately.

This quantity is conceptually different from standard posterior inclusion probabilities in Bayesian variable selection \citep{george1993variable,barbieri2004optimal}, which assign a single binary inclusion indicator to each predictor. SCP instead offers a summarization of the spatial significance map, which describes \emph{where in space} a predictor matters. We deem a predictor ``spatially informative" if \(\text{SCP}_j > 0.5\), meaning it has a credibly nonzero effect in more than half of the grid locations.

Importantly, SCP captures the spatial \emph{coverage} of significant effects, not the \emph{degree} of spatial variation. A predictor with a constant, nonzero effect everywhere will have \(\text{SCP}_j \approx 1\) despite no spatial heterogeneity, whereas a predictor that is strongly positive in one subregion and has no effect in other regions might have a low \(\text{SCP}_j\). Thus, SCP tells us whether a predictor has a meaningful effect on the response somewhere (and how widespread that effect is), rather than whether its effect changes across space.

To complement SCP, we also evaluate pointwise F1 scores and false positive rate (FPR) in simulation studies where the true coefficient surfaces are known. At each grid point \({\boldsymbol s}_g\), we treat \(\beta_j({\boldsymbol s}_g) \neq 0\) as a true signal and classify it as a detected signal if the credible interval excludes zero. For predictors with a true signal, the F1 score summarizes the trade-off between precision (the fraction of detected signal points that are truly nonzero) and recall (the fraction of true signal points correctly detected). For predictors that are truly null, the FPR quantifies the fraction of locations incorrectly flagged as significant. Together, SCP, F1, and FPR provide a detailed assessment of spatial variable selection performance in simulations.

\section{Simulation Study}\label{sec:simulation}

\subsection{Simulation Design}\label{subsec:sim_design}

We conduct simulation studies to evaluate the performance of the proposed Bayesian spatial group lasso (BSGL) and compare it with two established SVC methods: GGP-GAM \citep{comber2024multiscale} and MGWR \citep{fotheringham2017multiscale}. %We also compare it with non-spatial variable selection methods, lasso and adaptive lasso \citep{zou2006adaptive}.
Simulations are conducted across various configurations with sample sizes $n \in \{1000, 2000, 5000, 10000\}$ and number of predictors $m \in \{5, 7, 10\}$. For each configuration,  spatial coordinates ${\boldsymbol z}_i = (u_i, v_i)$ are uniformly sampled from the domain $[0, 20] \times [0, 20]$. Following the simulation design of \cite{comber2024multiscale}, predictor variables $x_{j}({\boldsymbol z}_i)$ are generated from standardized normal distributions and then rescaled to the interval $[0,1]$. The idiosyncratic errors are simulated following $\varepsilon_i \stackrel{\text{iid}}{\sim} N(0, \sigma^2)$ with $\sigma^2 = 0.1$.

% All simulations were conducted on a Windows workstation with 32 CPU cores (x86-64 architecture) and 64GB RAM, using R version 4.3.1. Computation times are reported for single-threaded MCMC execution to ensure fair comparison across methods. The BSGL implementation supports parallel processing across multiple chains for convergence diagnostics, though timing comparisons use single-threaded runs.

For each simulation configuration, the first three variables $x_1, x_2, x_3$ are designated as signal variables with spatially varying coefficients:
\begin{subequations}
\begin{align}
\beta_1(u, v) &= 20 \cos\left(\frac{\pi u}{20}\right) \cos\left(\frac{\pi v}{20}\right), \\
\beta_2(u, v) &= 18 \cos\left(\frac{\pi u}{18}\right) \sin\left(\frac{\pi v}{18}\right), \\
\beta_3(u, v) &= 20 \exp\left(-\frac{(u-10)^2 + (v-10)^2}{50}\right).
\end{align}
\end{subequations}

These coefficient surfaces exhibit diverse spatial patterns: $\beta_1$ displays a checkerboard pattern, $\beta_2$ shows oscillatory structure, and $\beta_3$ presents a smooth Gaussian peak. The remaining variables are unstructured noise variables with zero coefficients ($\beta_j({\boldsymbol z}) = 0$ for $j = 4, \ldots, m$ and all ${\boldsymbol z}$), creating a sparse structure ideal for evaluating variable selection performance.

% \begin{figure}[!t]
%     \centering
%     \includegraphics[width=\textwidth]{pic/beta_functions_cowplot.png}
%     \caption{True coefficient surfaces for the first three variables. Each panel shows the spatial pattern with colors representing coefficient magnitudes.}
%     \label{fig:beta_functions}
% \end{figure}

We employ B-spline basis functions to represent the spatial coefficients. Following \citet{bai2019fast}, the number of basis functions $L$ is selected from a grid consisting of $\{16, 25, 36, 49\}$, corresponding to $4 \times 4$, $5 \times 5$, $6 \times 6$, and $7 \times 7$ tensor product B-spline bases with equal resolution in both spatial dimensions. The choice of optimal $L$ is made through 5-fold cross-validation. Hyperparameters for the group lasso prior $\lambda^2 \sim \text{Ga}(a_{\lambda}, b_{\lambda})$ are also selected via cross-validation \citep{wood2016smoothing} from a grid where $a_{\lambda} \in \{15, 30, 35, 40, 45\}$ and $b_{\lambda} \in \{0.01, 0.1, 1\}$. Cross-validation evaluates all combinations using 1000 MCMC iterations with 200 warm-up samples for computational efficiency, selecting the configuration that minimizes the average mean squared prediction error $\text{MSPE} = \frac{1}{n_I}\sum_{i \in I} (y({\boldsymbol z}_i) - \widehat{y({\boldsymbol z}_i)})^2$, where $I$ denotes the test set indices and $\widehat{y({\boldsymbol z}_i)}$ is the predicted response. Section 2 in the supplementary material offers more discussion on this.

The final model with the optimal choices of $L, a_{\lambda}, b_{\lambda}$ are fitted using four independent Markov chains to ensure reliable inference. Each chain runs for 5000 
iterations with the first 500 iterations discarded as warm-up, yielding 4,500 posterior samples for inference. Convergence is assessed using the Gelman-Rubin statistic $\hat{R}$ \citep{gelman1992inference}, effective sample size, and visual inspection of trace plots. In all simulation 
scenarios, we achieved $\hat{R} < 1.01$ for all parameters, and trace plots for key hyperparameters in a representative simulation (see Section 3 of the supplementary material) demonstrate good mixing across all chains. 

Performance is evaluated through multiple metrics. MSPE on a 20\% held-out test set measures overall predictive accuracy. MSE$_1$ measures the mean squared error between estimated and true coefficient surfaces for the first three variables with true non-zero effects, defined as $\text{MSE}_1 = \frac{1}{3}\sum_{j=1}^{3} \frac{1}{n}\sum_{i=1}^{n} (\widehat{\beta}_j(z_i) - \beta_j(z_i))^2$. MSE$_0$ measures estimation accuracy for variables with true zero coefficients, defined as $\text{MSE}_0 = \frac{1}{m-3}\sum_{j=4}^{m} \frac{1}{n}\sum_{i=1}^{n} \widehat{\beta}_j^2(z_i)$.  Variable selection performance is evaluated through spatial coverage probabilities $\text{SCP}_j$.

% Coverage rate measures the proportion of test observations within 95\% prediction intervals for uncertainty quantification.
The proposed BSGL is compared with GGP-GAM implemented using the \texttt{mgcv} R package and MGWR to establish relative performance advantages.%, and with baselines lasso and adaptive lasso \citep{zou2006adaptive} implemented using \texttt{glmnet} \citep{friedman2010regularization}.

\subsection{Simulation Results}\label{subsec:sim_results}
\subsubsection{Coefficient Estimation and Prediction Performance}
We compare BSGL with GGP-GAM and MGWR to illustrate both the importance of modeling spatial heterogeneity and the benefits of integrated variable selection. Table~\ref{tab:simresults} summarizes results for BSGL and the two competing spatially varying coefficient methods, which differ primarily in how (or whether) they perform spatially varying coefficient estimation. All approaches are assessed under identical simulation settings across multiple sample sizes and numbers of predictors.

BSGL delivers prediction accuracy that is competitive with or better than GGP-GAM in all configurations. For smaller samples ($n = 1000, 2000$), GGP-GAM attains slightly lower MSPE in most predictor settings. As the sample size grows, however, BSGL’s performance improves more rapidly. At $n = 5000$, BSGL yields lower MSPE than GGP-GAM when $m = 10$ and comparable MSPE when $m = 5$ or $7$. The advantage becomes clear at $n = 10{,}000$, where BSGL consistently attains lower MSPE across all predictor dimensions. Both BSGL and GGP-GAM substantially outperform MGWR in every setting: MGWR’s MSPE is roughly two to four times larger at small sample sizes and about 30--50\% larger at the largest sample size.

The largest gains for BSGL appear in estimation accuracy for spatially varying signal coefficients, measured by $\text{MSE}_1$. At $m = 1000$, GGP-GAM has some advantage, but this reverses at larger samples. For example, with $n = 5000$ and $m = 5$, BSGL attains $\text{MSE}_1 = 0.031$ versus 0.069 for GGP-GAM, an improvement of about 55\%. With $m = 7$ and the same sample size, BSGL achieves 0.038 compared to 0.063 for GGP-GAM (roughly a 40\% reduction), and similar patterns hold for $m = 10$. At $n = 10{,}000$, BSGL maintains a roughly two- to three-fold reduction in $\text{MSE}_1$ across all predictor dimensions. These gains reflect the BSGL shrinkage prior, which concentrates estimation effort on true signal coefficients while strongly shrinking noise coefficients.

For noise coefficients, assessed via $\text{MSE}_0$, the comparison is more nuanced. At smaller sample sizes and moderate predictor counts ($m = 5, 7$), GGP-GAM typically attains lower $\text{MSE}_0$, consistent with its smooth estimation for all coefficients. As sample size increases, BSGL becomes competitive and often superior in shrinking noise effects. For instance, at $n = 10{,}000$ with $m = 10$, BSGL achieves the lowest $\text{MSE}_0$ among the three methods. In these larger-sample, higher-dimensional settings, the BSGL shrinkage prior more effectively identifies noise variables and shrinks their coefficients toward zero.
\begin{table}[!htbp]
\centering
\caption{Performance comparison between BSGL, GGP-GAM, and MGWR across sample sizes ($n$) and number of predictors ($m$). We show mean squared prediction error (MSPE), mean squared error for signal variables MSE$_1$, and mean squared error for noise variables MSE$_0$. BSGL shows superior performance as $n$ and $m$ increase.}
\label{tab:simresults}
\small
\begin{tabular}{@{}cccccc@{}}
\toprule
n & m & \textbf{Method} & \textbf{MSPE} & \textbf{MSE$_1$} & \textbf{MSE$_0$} \\
\midrule
1000 & 5 & BSGL & \textbf{0.115} & 0.127 & 0.032 \\
     &   & GGP-GAM & 0.120 & \textbf{0.119} & \textbf{0.007} \\
     &   & MGWR & 0.446 & 0.901 & 0.015 \\
\midrule
1000 & 7 & BSGL & 0.142 & 0.171 & 0.062 \\
     &   & GGP-GAM & \textbf{0.131} & \textbf{0.106} & 0.035 \\
     &   & MGWR & 0.469 & 0.852 & \textbf{0.016} \\
\midrule
1000 & 10 & BSGL & 0.163 & 0.268 & 0.072 \\
     &    & GGP-GAM & \textbf{0.153} & \textbf{0.144} & 0.051 \\
     &    & MGWR & 0.455 & 0.967 & \textbf{0.030} \\
\midrule
2000 & 5 & BSGL & 0.108 & 0.098 & 0.033 \\
     &   & GGP-GAM & \textbf{0.105} & \textbf{0.087} & \textbf{0.004} \\
     &   & MGWR & 0.291 & 0.514 & 0.006 \\
\midrule
2000 & 7 & BSGL & 0.118 & 0.094 & 0.045 \\
     &   & GGP-GAM & \textbf{0.114} & \textbf{0.090} & \textbf{0.017}\\
     &   & MGWR & 0.280 & 0.512 & 0.065 \\
\midrule
2000 & 10 & BSGL & \textbf{0.114} & \textbf{0.083} & 0.029 \\
     &    & GGP-GAM & 0.116 & 0.089 & 0.028 \\
     &    & MGWR & 0.243 & 0.544 & \textbf{0.014} \\
\midrule
5000 & 5 & BSGL & 0.117 & \textbf{0.031} & 0.033 \\
     &   & GGP-GAM & \textbf{0.115} & 0.069 & \textbf{0.003} \\
     &   & MGWR & 0.177 & 0.215 & 0.025\\
\midrule
5000 & 7 & BSGL & 0.115 & \textbf{0.038} & \textbf{0.011} \\
     &   & GGP-GAM & \textbf{0.113} & 0.063 & 0.014 \\
     &   & MGWR & 0.168 & 0.220 & 0.013 \\
\midrule
5000 & 10 & BSGL & \textbf{0.107} & \textbf{0.036} & \textbf{0.013} \\
     &    & GGP-GAM & 0.108 & 0.054 & 0.016 \\
     &    & MGWR & 0.160 & 0.222 & 0.015 \\
\midrule
10000 & 5 & BSGL & \textbf{0.105} & \textbf{0.019} & 0.008 \\
      &   & GGP-GAM & 0.107 & 0.057 & \textbf{0.003} \\
      &   & MGWR & 0.138 & 0.128 & 0.004 \\
\midrule
10000 & 7 & BSGL & \textbf{0.105} & \textbf{0.022} & 0.010 \\
      &   & GGP-GAM & 0.110 & 0.057 & 0.008 \\
      &   & MGWR & 0.137 & 0.128 & \textbf{0.006} \\
\midrule
10000 & 10 & BSGL & \textbf{0.100} & \textbf{0.020} & \textbf{0.008} \\
      &    & GGP-GAM & 0.104 & 0.049 & 0.012 \\
      &    & MGWR & 0.131 & 0.127 & 0.001 \\
\bottomrule
\end{tabular}
%\begin{tablenotes}
%\small
%\item Note: MSPE = mean squared prediction error on test set; MSE$_1$ = mean squared error for signal coefficients $\beta_1$--$\beta_3$; MSE$_0$ = mean squared error for noise coefficients $\beta_4$--$\beta_{10}$. Bold values indicate best performance in each scenario.
%\end{tablenotes}
\end{table}

\subsubsection{Spatial Variable Selection}
We show results for spatial variable selection for the most challenging case of $m=10$ predictors. The conclusions for $m=5, 7$ predictors are similar. For $m=10$ predictors, Figure~\ref{fig:beta_recon} presents visual comparisons of estimated spatial coefficients for all signal variables $\beta_1({\boldsymbol z}),\beta_2({\boldsymbol z}),\beta_3({\boldsymbol z})$ and a representative noise variable $\beta_5({\boldsymbol z})$ for $n = 1000$.  For signal variables with spatial effects, all three methods successfully capture the main spatial patterns. BSGL produces smooth and interpretable coefficient surfaces that accurately reflect the underlying spatial heterogeneity, with reconstruction quality comparable to GGP-GAM.

\begin{figure}[!tbp]
    \centering
    \includegraphics[width=\textwidth]{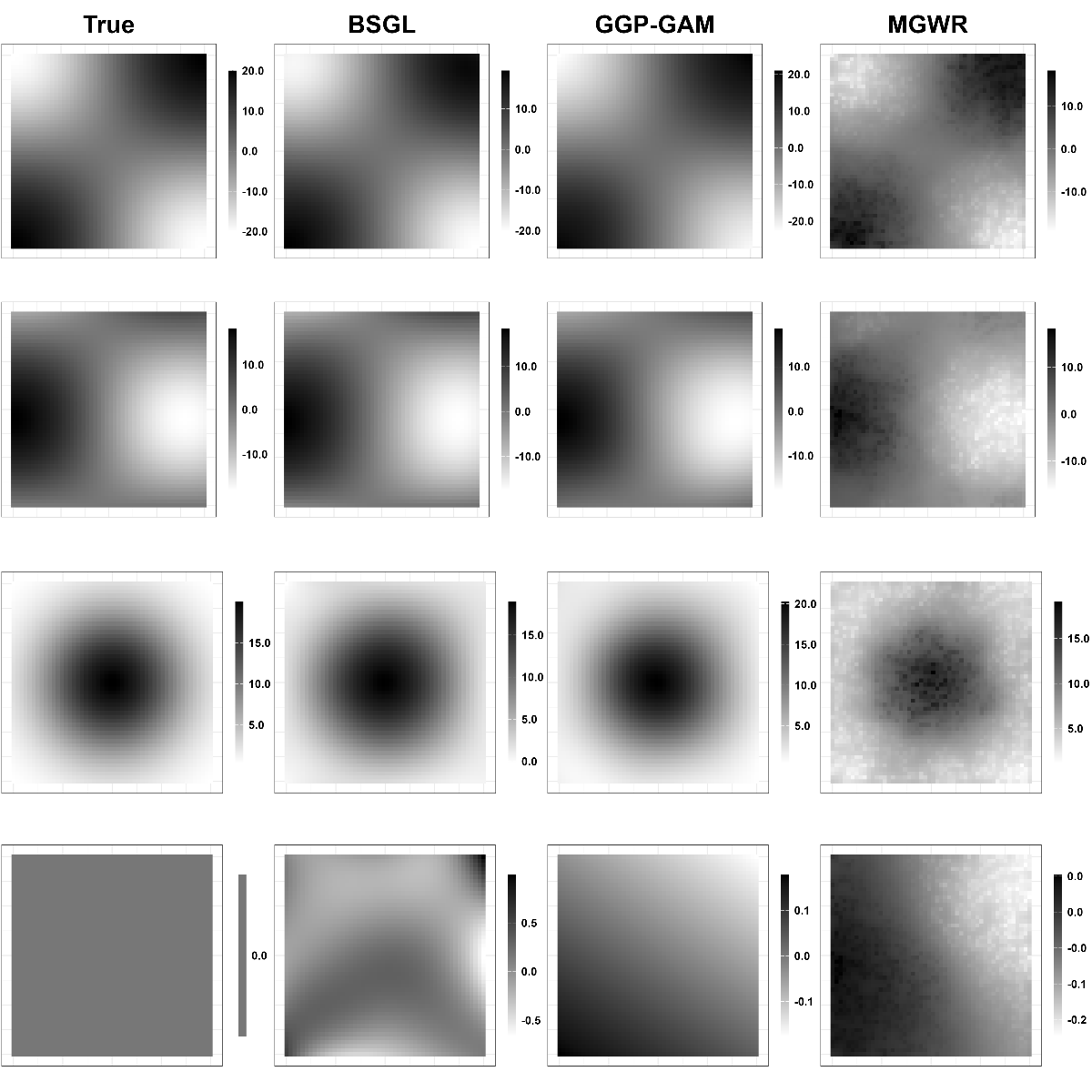}
    
    \caption{Spatial coefficient estimates for signal variables $\beta_1$-$\beta_3$ as shown in top three panels and noise variable $\beta_5$ in bottom panel at $n=1000$, $m=5$. All methods capture the true spatial patterns for signal variables, with BSGL providing sharper boundary estimates. For the zero predictor $\beta_5$, BSGL correctly shrinks coefficients to near-zero across the domain while GGP-GAM and MGWR fit spurious spatial patterns.}
    \label{fig:beta_recon}
\end{figure}

The critical difference emerges when examining noise variables. While the true $\beta_5({\boldsymbol z})$ surface is zero everywhere, GGP-GAM and MGWR produce surfaces with artificial spatial gradients and spurious local variations. In contrast, BSGL correctly recovers a surface close to zero across the entire domain, demonstrating the effectiveness of the group lasso prior in identifying predictors with negligible effects.

We further evaluate variable selection using SCP, F1 scores, and FPR for simulation experiments.  Table~\ref{tab:scp_f1_rates} presents these metrics across sample sizes using 95\% credible intervals. Signal variables achieve high SCP values ranging from 0.86 to 0.99 and F1 scores between 0.93 and 0.96, indicating strong detection of nonzero regions of the coefficient surfaces with minimal false negatives. Noise variables maintain low SCP values, predominantly below 0.05, with F1 scores of zero in all cases, correctly identified as having negligible effects. FPR remains below 0.05 for all noise variables, with most at zero, demonstrating good specificity. Figure~\ref{fig:pip_maps} visualizes this separation through spatial significance maps at sample size $5000$. Signal variables in the top panel display consistent spatial regions where 95\% credible intervals exclude zero, shown as solid points covering most of the spatial domain with SCP exceeding 0.90. Noise variables in the bottom panel show predominantly hollow points with SCP below 0.05, consistent with their true zero coefficients in the data setting. The clear separation between signal and noise variables enables reliable variable selection across all sample sizes. It is also important to note that the separation of SCP values between signal and noise predictors increases as the sample size increases. We also argue that the variable selection is acceptably robust for moderate perturbation of the significance level $\alpha$. To demonstrate this, additional SCP results across credible interval levels from 70\% to 95\% appear in supplementary materials Section 4.1. Unlike GGP-GAM and MGWR which fit spatial surfaces for all covariates regardless of their  spatial behavior, BSGL achieves complete variable identification while controlling false discoveries.

\begin{table}[!htbp]
\centering
\caption{Performance metrics across sample sizes for signal variables $\beta_1$, $\beta_2$, $\beta_3$ and noise variables $\beta_4$--$\beta_{10}$ using 95\% credible intervals.}
\label{tab:scp_f1_rates}
\small
\begin{tabular}{@{}cc|ccc|ccccccc@{}}
\toprule
$n$ & Metric & $\beta_1$ & $\beta_2$ & $\beta_3$ & $\beta_4$ & $\beta_5$ & $\beta_6$ & $\beta_7$ & $\beta_8$ & $\beta_9$ & $\beta_{10}$ \\
\midrule
\multirow{3}{*}{1000} 
& SCP & 0.92 & 0.86 & 0.91 & 0.00 & 0.00 & 0.01 & 0.00 & 0.02 & 0.05 & 0.01 \\
& F1 & 0.96 & 0.93 & 0.95 & 0.00 & 0.00 & 0.00 & 0.00 & 0.00 & 0.00 & 0.00 \\
& FPR & 0.00 & 0.00 & 0.00 & 0.00 & 0.00 & 0.01 & 0.00 & 0.02 & 0.05 & 0.01 \\
\cmidrule(lr){1-12}
\multirow{3}{*}{2000} 
& SCP & 0.95 & 0.90 & 0.95 & 0.00 & 0.02 & 0.00 & 0.00 & 0.01 & 0.00 & 0.00 \\
& F1 & 0.97 & 0.96 & 0.98 & 0.00 & 0.00 & 0.00 & 0.00 & 0.00 & 0.00 & 0.00 \\
& FPR & 0.00 & 0.00 & 0.00 & 0.00 & 0.03 & 0.00 & 0.00 & 0.01 & 0.00 & 0.00 \\
\cmidrule(lr){1-12}
\multirow{3}{*}{5000} 
& SCP & 0.97 & 0.94 & 0.98 & 0.00 & 0.00 & 0.03 & 0.00 & 0.00 & 0.02 & 0.00 \\
& F1 & 0.99 & 0.98 & 0.99 & 0.00 & 0.00 & 0.00 & 0.00 & 0.00 & 0.00 & 0.00 \\
& FPR & 0.00 & 0.00 & 0.00 & 0.00 & 0.01 & 0.02 & 0.00 & 0.00 & 0.02 & 0.01 \\
\cmidrule(lr){1-12}
\multirow{3}{*}{10000} 
& SCP & 0.99 & 0.94 & 0.99 & 0.00 & 0.02 & 0.02 & 0.05 & 0.00 & 0.00 & 0.06 \\
& F1 & 0.99 & 0.98 & 1.00 & 0.00 & 0.00 & 0.00 & 0.00 & 0.00 & 0.00 & 0.00 \\
& FPR & 0.00 & 0.00 & 0.00 & 0.00 & 0.03 & 0.02 & 0.04 & 0.00 & 0.00 & 0.05 \\
\bottomrule
\end{tabular}
\end{table}

\begin{figure}[!htbp]
    \centering
    \includegraphics[width=\textwidth]{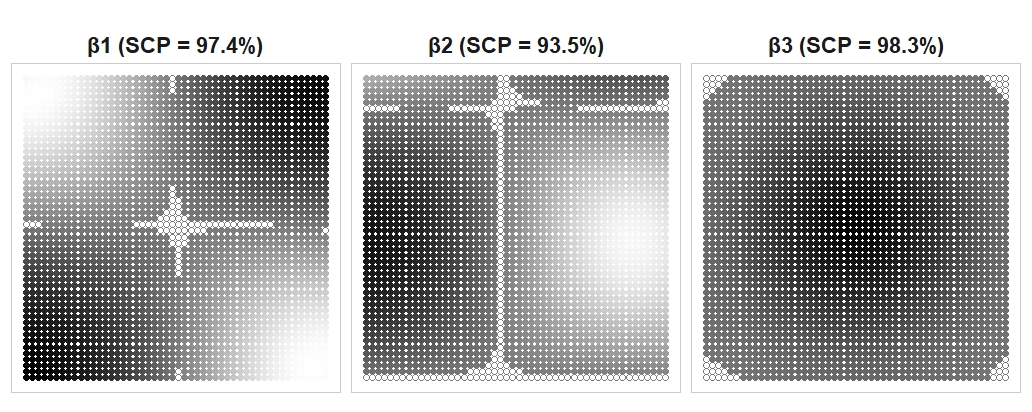}
    \vspace{-2mm} \includegraphics[width=\textwidth]{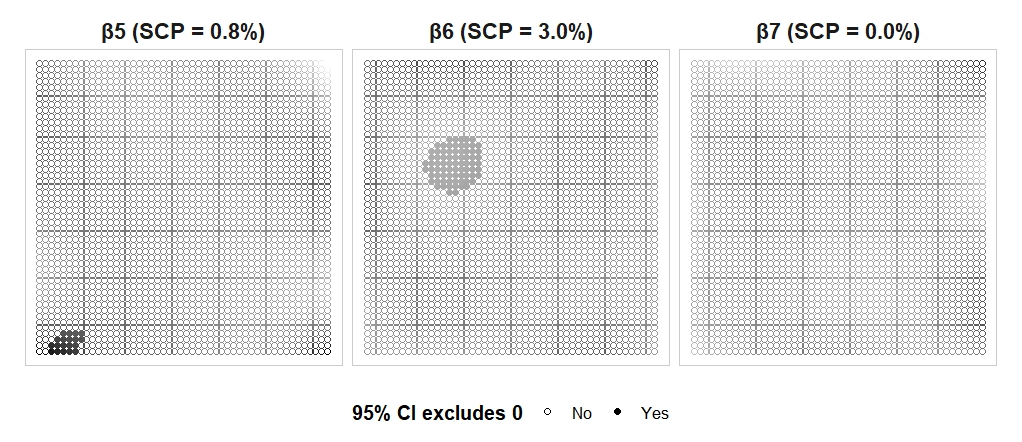}
    \caption{Spatial significance maps with 95\% credible intervals. Solid points indicate locations where credible intervals exclude zero. Signal variables in the top panel achieve SCP exceeding 0.90. Noise variables in the bottom panel achieve SCP below 0.03.}
    \label{fig:pip_maps}
\end{figure}

We further evaluate the method's behavior for spatially constant but non-zero coefficients by setting $\beta_4({\boldsymbol z}) = c$, for all locations. Detection depends on signal strength: weak constants with $c = 0.1$ achieve SCP of 0.00, moderate constants with $c = 0.5$ achieve SCP of 0.13, while strong constants with $c = 1.0$ and $c = 10.0$ achieve SCP of 0.78 and 1.00, respectively. This reflects the method's design to prioritize spatial variation, applying stronger shrinkage when spatial structure is minimal. Detailed analysis appears in the supplementary materials Section 4.2.

 All simulations were conducted on a Windows workstation with 32 CPU cores using x86-64 architecture and 64 GB RAM, using R version 4.3.1. At $n = 10{,}000$ with $m = 10$ predictors, MGWR requires 1,591 minutes, including cross-validation and model fitting, while BSGL completes in 91 minutes with 83 minutes for cross-validation and 8 minutes for final model estimation. The BSGL implementation supports parallel processing across multiple cores. However, the computational efficiency decays when more basis functions are used. We offer strategies to scale our approach in Section~\ref{sec:discussion}.

\section{Application to Vegetation Data}\label{sec:application}

The western United States has experienced unprecedented ecological transformation in recent decades, with intensifying drought stress, record-breaking wildfire seasons, and rapid shifts in vegetation composition \citep{williams2020large,abatzoglou2016impact}. Understanding which environmental factors drive vegetation and whether these relationships exhibit \color{black} non-zero spatial effects \color{black} is essential for predicting ecosystem responses and developing targeted management strategies.

We analyze MODIS Enhanced Vegetation Index data from the western United States, specifically sinusoidal grid tile h08v05 covering approximately 30°N--40°N latitude and 104°W--130°W longitude. This region encompasses extreme environmental gradients from coastal temperate rainforests receiving over 2000 mm annual precipitation to interior deserts with less than 200 mm, supporting diverse ecosystems including Pacific coastal forests, Sierra Nevada montane systems, Great Basin shrublands, and Mojave Desert communities. The original dataset contains 1.02 million observations with 28 variables collected during 2016.

Given that the full dataset presents computational challenges due to increased run time, we randomly subsample 10,000 observations to ensure computational feasibility while maintaining spatial representativeness across the study region. This subsample size is sufficient to capture the main spatial patterns, given the large geographic extent and relatively smooth spatial processes characterizing vegetation-environment relationships at landscape scales. Sensitivity analysis comparing model estimates across sample sizes from 1,000 to 15,000 observations demonstrates inferential stability at moderate to large sample sizes. The 10,000 observation subsample achieves stable spatial coverage probabilities and predictive accuracy comparable to larger samples, validating our sample size choice for computational efficiency while maintaining statistical reliability. A detailed discussion on this topic is offered in Section 5.1 of the supplementary material.

The response variable is the Enhanced Vegetation Index, which provides improved sensitivity over dense vegetation and enhanced atmospheric correction relative to traditional indices \citep{huete2002overview}. We select 10 environmental predictors representing different aspects of vegetation-environment relationships: spectral reflectance bands (red, near-infrared (NIR), blue, mid-infrared (MIR)), ecosystem productivity measures (Gross Primary Productivity (GPP), Latent Heat Flux (LE)), observation geometry variables (view zenith angle, sun zenith angle, relative azimuth angle), and land surface characteristics (land cover type). We apply the transformation $\log(\text{EVI} + 1)$ to improve normality and split the data into 80\% training and 20\% testing sets. All predictors are standardized to the $[0,1]$ range.

Model hyperparameters are selected via 5-fold cross-validation over $L \in \{16, 25, 36\}$, $a_\lambda \in \{5, 7, 10, 20, 30\}$, $b_\lambda \in \{0.2, 0.3, 0.5, 1\}$. The final model uses $L = 25$ basis functions with $a_\lambda = 20$, $b_\lambda = 0.5$, estimated using 8,000 MCMC iterations with 1,000 warm-up samples across four independent chains. Convergence diagnostics confirm satisfactory mixing with $\hat{R} < 1.02$ for all parameters.

BSGL achieves a mean squared prediction error of $6.34 \times 10^{-5}$ on the held-out test set, with GGP-GAM and MGWR achieving $6.81 \times 10^{-5}$ and $1.50 \times 10^{-4}$, respectively. The model provides a predictive coverage of 96.55\% for 95\% prediction intervals, indicating accurate predictions and well-calibrated uncertainty quantification.
Figure~\ref{fig:evi_spatial_comparison} demonstrates the model's ability to capture complex spatial patterns in vegetation indices across the study region. The predicted values closely track observed patterns, successfully reproducing the transition from high-productivity coastal and montane systems to low-productivity interior deserts. Figure~\ref{fig:evi_spatial_comparison} shows prediction errors being small, scattered randomly across the study region. Spatial autocorrelation analysis yields Moran's $I = 0.03$ with $p$-value $= 0.42$, indicating no systematic spatial structure. Detailed residual diagnostics are provided in the supplementary materials Section 5.2. 

\begin{figure}[!t]
\centering
\includegraphics[width=\textwidth]{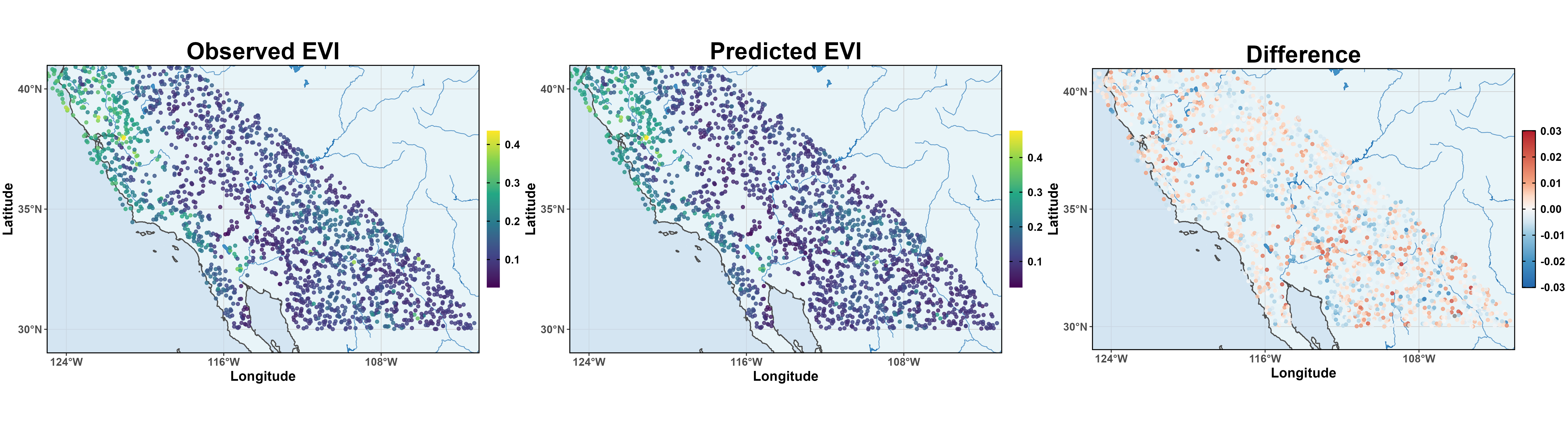}
\caption{Spatial distribution of Enhanced Vegetation Index across the test region showing observed values, model predictions, and residuals.}
\label{fig:evi_spatial_comparison}
\end{figure}

\color{black}
To evaluate variable selection capabilities, we compare BSGL with lasso and adaptive lasso. Table~\ref{tab:method_comparison} summarizes variable selection results across methods. Standard lasso selects 9 of 10 predictors, demonstrating limited selectivity by retaining nearly all variables. Adaptive lasso improves variable selection by identifying 6 variables: red reflectance, NIR reflectance, blue reflectance, MIR reflectance, GPP, and LE. We note that while GGP-GAM excels at modeling spatially varying relationships \citep{comber2024multiscale}, it is fundamentally not designed for variable selection tasks.
\color{black}

\begin{table}[!t]
\centering
\caption{Variable selection results across modeling approaches.}
\label{tab:method_comparison}
\small
\begin{tabular}{@{}lcccc@{}}
\toprule
\textbf{Variable} & \textbf{lasso} & \textbf{Adaptive lasso} & \textbf{BSGL} \\
 & $|\beta|$ & $|\beta|$ & SCP \\
\midrule
Red reflectance & 0.326 & 0.331 & 0.98 \\
NIR reflectance & 0.349 & 0.352 & 0.98 \\
Blue reflectance & 0.031 & 0.033 & 0.76 \\
MIR reflectance & 0.015 & 0.013 & 0.13 \\
GPP & 0.059 & 0.061 & 0.93 \\
LE & 0.007 & 0.003 & 0.83 \\
View zenith angle & 0.000 & 0.000 & 0.27 \\
Sun zenith angle & 0.002 & 0.000 & 0.80 \\
Relative azimuth angle & 0.001 & 0.000 & 0.60 \\
LC Type4 & 0.002 & 0.000 & 0.09 \\
\bottomrule
\multicolumn{4}{l}{\footnotesize 0.000 indicates variable was excluded from lasso models.} \\
\end{tabular}
\end{table}

In contrast, BSGL enables selective identification of predictors that exhibit spatially extensive and credibly nonzero effects by utilizing SCP. Three variables display strong evidence of nonzero spatial effects, with SCP values exceeding 0.90: NIR reflectance (SCP $= 0.98$), Red reflectance (SCP $= 0.98$), and GPP (SCP $= 0.93$). Four variables show moderate spatial significance: LE (SCP $= 0.83$), Sun zenith angle (SCP $= 0.80$), Blue reflectance (SCP $= 0.76$), and Relative azimuth angle (SCP $= 0.60$). The remaining three variables, View zenith angle, MIR reflectance, and LC Type4, have coefficients that are not credibly different from zero across most spatial locations, with SCP values at or below 0.50.

% 2x2 SCP version, please uncomment below.
%\begin{figure}[!t]
%\centering
%\includegraphics[width=\textwidth]{pic/scp_maps_2x2.png}
%\caption{SCP maps contrasting predictor variables with highest (NIR reflectance, red reflectance) and lowest (MIR reflectance, LC Type4) spatial significance. Dark blue indicates locations where 95\% credible intervals exclude zero; light blue shows intervals including zero.}
%\label{fig:scp_maps}
%\end{figure}

%\color{green}
Further, Figure~\ref{fig:scp_maps} contrasts spatial significance patterns between predictors with the strongest and weakest spatial coverage. The top row shows NIR and red reflectance, which exhibit nearly complete spatial significance across the study region. Similarly, the third row shows Gross primary productivity (GPP) which also exhibit spatial significance almost in the entire domain. Their credible intervals exclude zero throughout most locations, visualized in dark blue. In contrast, the second row and the last row display MIR reflectance and land cover type (LC Type4), respectively, where spatial significance is minimal. These variables show credible intervals containing zero across nearly the entire domain, appearing in light blue. These patterns demonstrate BSGL's ability to identify which environmental predictors exhibit spatially extensive effects versus those with spatially limited or negligible relationships with vegetation indices.
%\color{black}

% 5x2 SCP version, please uncomment below.

 \begin{figure}[!htbp]
 \centering
 \includegraphics[width=0.7\textwidth]{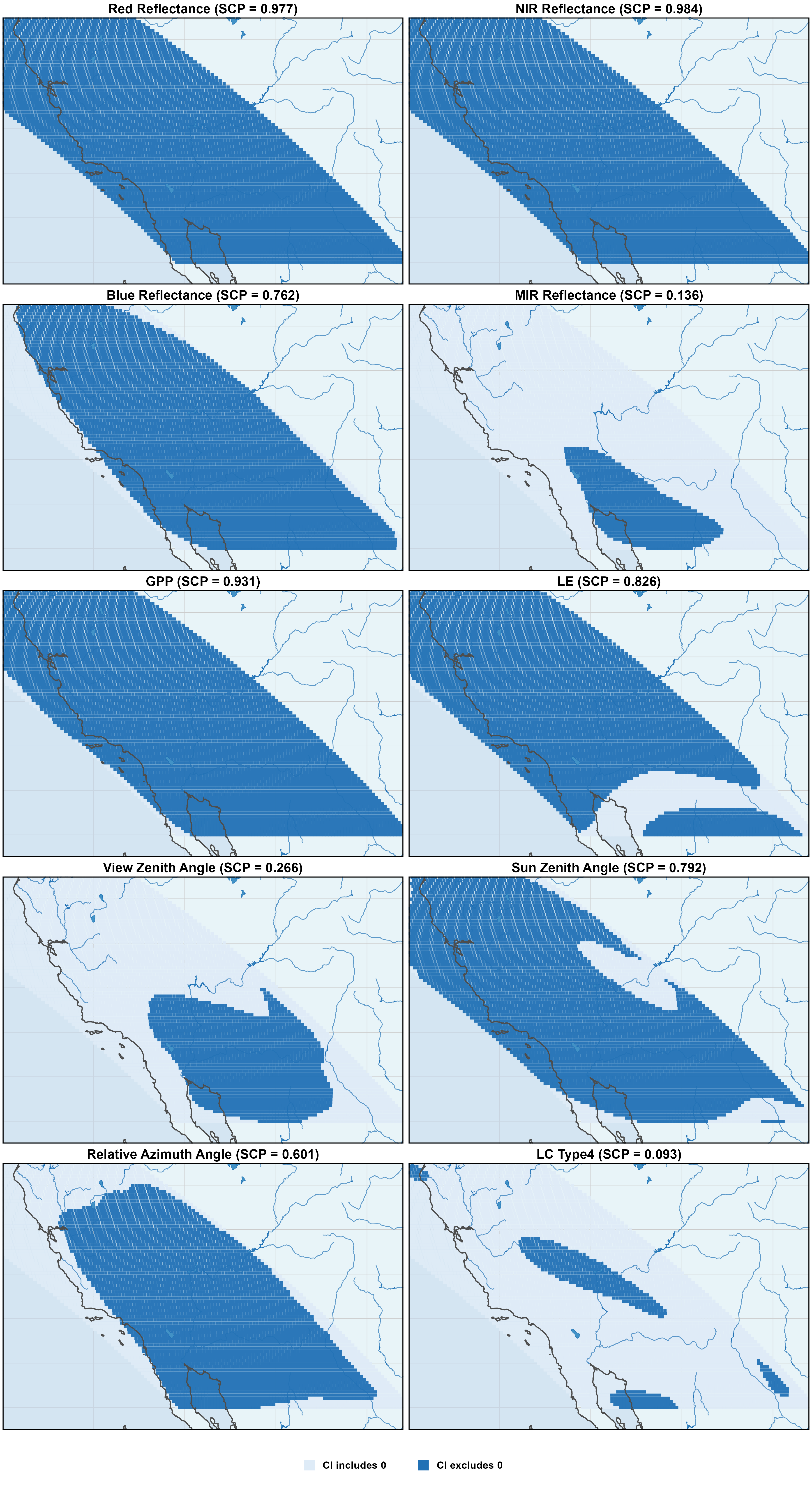}
 \caption{Spatial coverage probability maps for predictor variables. Dark blue regions indicate locations where 95\% credible intervals exclude zero. Spectral reflectance (red, NIR, blue) and productivity variables (GPP, LE) exhibit high spatial significance across the study region, while observation geometry and land cover variables show limited or spatially localized patterns.}
 \label{fig:scp_maps}
 \end{figure}

The empirical findings offer valuable insights into vegetation dynamics. The pronounced spatial significance of Gross primary productivity (GPP) likely reflect moisture gradients and growing season length, with stronger positive GPP-EVI relationships in regions where water availability supports sustained photosynthetic activity. This gradient pattern has direct implications for predicting vegetation responses to drought intensification under climate change, as the strength of productivity-vegetation coupling varies systematically across the landscape.
Sun zenith angle and latent heat flux exhibit credibly nonzero effects over considerable spatial extents. Sun zenith angle shows stronger effects in mountainous terrain where topographic shading and illumination geometry interactions vary substantially, corresponding closely to terrain complexity in the Sierra Nevada and coastal ranges compared to interior basins. Latent heat flux spatial patterns reflect regional differences in evapotranspiration across moisture and vegetation gradients.

Variables with low SCP provide equally important insights. MIR reflectance, view zenith angle, and LC Type4 show SCP values at or near zero, indicating coefficients not significantly different from zero across most spatial locations. For MIR reflectance, this suggests consistent relationships between canopy moisture content and spectral properties regardless of spatial location. The near-zero spatial coverage for land cover type indicates that after accounting for spectral and productivity variables, discrete land cover categories contribute little additional explanatory signal, as indicated by near-zero SCP. 

BSGL results suggest focusing interpretation and modeling effort on NIR reflectance, red reflectance, and gross primary productivity, which show spatially extensive, credibly nonzero effects with SCP exceeding 0.90, along with latent heat flux and sun zenith angle that demonstrate substantial spatial significance. Variables with near-zero spatial coverage can be treated with simpler model specifications. Unlike standard GGP-GAM approaches that fit spatial surfaces for all variables, BSGL distinguishes which environmental drivers have spatially extensive, credibly nonzero effects, focusing resources on variables with practically nonnegligible associations and de-emphasizing variables with negligible effects.

\section{Discussion}\label{sec:discussion}

The proposed BSGL framework provides a rigorous and practically effective approach to variable selection in spatially varying coefficient models. By differentiating environmental drivers that exhibit spatially extensive, credibly nonzero effects from those with negligible influence, BSGL yields the focused insight necessary for environmental monitoring and management, insight that standard GWR, MGWR, and GGP-GAM methods often fail to deliver. Whereas these existing approaches typically estimate nonzero spatial surfaces for all predictors, BSGL explicitly separates signal from noise, enabling researchers to concentrate on genuinely influential variables and avoid overinterpreting weak or spurious spatial patterns. In addition, BSGL produces spatial significance maps for each predictor, indicating where within the domain an effect is statistically significant and where it is not.

In the application to MODIS vegetation data across the western United States, BSGL uncovers ecologically interpretable spatial heterogeneity in near-infrared reflectance, red reflectance, and gross primary productivity, with coefficient surfaces that vary systematically along moisture and elevation gradients. Latent heat flux and sun zenith angle also demonstrate substantial spatial significance in their effects on vegetation. These results are consistent with established vegetation science, capturing well-known transitions from coastal forests to interior deserts and reflecting variation in growing season length and water availability. Simultaneously, the method identifies variables such as mid-infrared reflectance, view zenith angle, and land cover type as lacking spatially significant effects, thereby streamlining model interpretation and sharpening monitoring designs. This dual role, both confirming theoretical expectations and indicating which predictors can be deprioritized, highlights the value of BSGL for both scientific inference and applied decision-making.

From a management and policy standpoint, the spatial patterns detected by BSGL have clear implications for climate adaptation and resource allocation. Areas where near-infrared and red reflectance display strong but opposing effects correspond to highly productive ecosystems that may be especially vulnerable to increasing drought stress. Spatial variation in the coupling between productivity and vegetation suggests that climate impacts will differ across environmental gradients, necessitating geographically targeted adaptation strategies. By pinpointing where vegetation–environment relationships are strongest and most spatially structured, BSGL helps prioritize locations for enhanced monitoring, conservation investments, and adaptive management actions.

Simulation results further support the robustness and generality of the method. BSGL reliably achieves near-perfect discrimination between signal predictors and noise predictors, with signal variables consistently exhibiting high SCP values, while noise variables remain close to zero, and this separation persists across a moderate range of credible interval levels, indicating stable variable selection that is not overly sensitive to the choice of significance threshold.

The BSGL framework also provides a flexible platform for future extensions. Incorporating spatiotemporal components would make it possible to examine how spatial patterns evolve under climate change, distinguishing stable from shifting relationships. Extending the framework to accommodate non-Gaussian responses would widen its applicability to species distributions, biodiversity counts, and other common ecological outcomes. While the present analysis focuses on samples up to $n\sim 10,000$, recent developments in distributed Bayesian computation \citep{guhaniyogi2022distributed, guhaniyogi2023distributed} and data sketching approaches \citep{guhaniyogi2021sketching, guhaniyogi2025bayesiandatasketchingvarying} could be integrated with BSGL to ensure scalability to much larger datasets. Beyond vegetation monitoring, the methodology is well suited to digital soil mapping, biodiversity modeling, ecosystem service valuation, and climate impact assessment—settings characterized by high-dimensional spatial data where determining which variables merit spatially explicit treatment is both difficult and crucial.

\section*{Supplementary Information}
Supplementary materials include detailed MCMC derivations, hyperparameter tuning results, convergence diagnostics, additional simulation results, and complete R code for implementing BSGL.

\section*{Acknowledgements}
We gratefully acknowledge Dr. Laura Baracaldo for providing the MODIS vegetation data used in the real data application. We thank the reviewers for their constructive comments that improved the manuscript.

\section*{Declarations}

\subsection*{Funding}
This research received no specific grant from any funding agency in the public, commercial, or not-for-profit sectors.

\subsection*{Conflict of Interest}
The authors declare no competing interests.

\subsection*{Data and Code Availability}
Simulated datasets and code to reproduce all analyses are available at \url{https://github.com/Qishi7/BSGL}.

\bibliographystyle{plainnat}
\bibliography{inference}

@article{scheel2013bayesian,
  title={A Bayesian hierarchical model with spatial variable selection: the effect of weather on insurance claims},
  author={Scheel, Ida and Ferkingstad, Egil and Frigessi, Arnoldo and Haug, Ola and Hinnerichsen, Mikkel and Meze-Hausken, Elisabeth},
  journal={Journal of the Royal Statistical Society Series C: Applied Statistics},
  volume={62},
  number={1},
  pages={85--100},
  year={2013},
  publisher={Oxford University Press}
}

@article{tucker1979red,
  title={Red and photographic infrared linear combinations for monitoring vegetation},
  author={Tucker, Compton J},
  journal={Remote sensing of Environment},
  volume={8},
  number={2},
  pages={127--150},
  year={1979},
  publisher={Elsevier}
}

@article{huete2002overview,
  title={Overview of the radiometric and biophysical performance of the MODIS vegetation indices},
  author={Huete, Alfredo and Didan, Kamel and Miura, Tomoaki and Rodriguez, E Patricia and Gao, Xiang and Ferreira, Laerte G},
  journal={Remote sensing of environment},
  volume={83},
  number={1-2},
  pages={195--213},
  year={2002},
  publisher={Elsevier}
}

@article{guhaniyogi2022distributed,
  title={Distributed Bayesian varying coefficient modeling using a {G}aussian process prior},
  author={Guhaniyogi, Rajarshi and Li, Cheng and Savitsky, Terrance D and Srivastava, Sanvesh},
  journal={Journal of machine learning research},
  volume={23},
  number={84},
  pages={1--59},
  year={2022}
}

@article{eilers1996flexible,
  title={Flexible smoothing with B-splines and penalties},
  author={Eilers, Paul HC and Marx, Brian D},
  journal={Statistical Science},
  volume={11},
  number={2},
  pages={89--121},
  year={1996}
}

@book{wahba1990spline,
  author    = {Wahba, Grace},
  title     = {Spline Models for Observational Data},
  year      = {1990},
  publisher = {Society for Industrial and Applied Mathematics},
  address   = {Philadelphia, PA},
  volume    = {59},
  isbn      = {9780898712445}
}

@article{guhaniyogi2021sketching,
  title={Sketching in Bayesian high dimensional regression with big data using {G}aussian scale mixture priors},
  author={Guhaniyogi, Rajarshi and Scheffler, Aaron},
  journal={arXiv preprint arXiv:2105.04795},
  year={2021}
}

@article{kammann2003geoadditive,
  title={Geoadditive models},
  author={Kammann, E. E. and Wand, M. P.},
  journal={Applied Statistics},
  volume={52},
  number={1},
  pages={1--18},
  year={2003}
}

@book{ruppert2003semiparametric,
  author    = {Ruppert, David and Wand, M. P. and Carroll, R. J.},
  title     = {Semiparametric Regression},
  year      = {2003},
  publisher = {Cambridge University Press},
  address   = {Cambridge, UK},
  series    = {Cambridge Series in Statistical and Probabilistic Mathematics}
}

@article{guhaniyogi2023distributed,
  title={Distributed Bayesian inference in massive spatial data},
  author={Guhaniyogi, Rajarshi and Li, Cheng and Savitsky, Terrance and Srivastava, Sanvesh},
  journal={Statistical science},
  volume={38},
  number={2},
  pages={262--284},
  year={2023},
  publisher={Institute of Mathematical Statistics}
}

@article{meyer2019importance,
  title={Importance of spatial predictor variable selection in machine learning applications--Moving from data reproduction to spatial prediction},
  author={Meyer, Hanna and Reudenbach, Christoph and W{\"o}llauer, Stephan and Nauss, Thomas},
  journal={Ecological Modelling},
  volume={411},
  pages={108815},
  year={2019},
  publisher={Elsevier}
}

@article{smith2003assessing,
  title={Assessing brain activity through spatial Bayesian variable selection},
  author={Smith, Michael and P{\"u}tz, Benno and Auer, Dorothee and Fahrmeir, Ludwig},
  journal={NeuroImage},
  volume={20},
  number={2},
  pages={802--815},
  year={2003},
  publisher={Elsevier}
}

@article{goldsmith2014smooth,
  title={Smooth scalar-on-image regression via spatial Bayesian variable selection},
  author={Goldsmith, Jeff and Huang, Lei and Crainiceanu, Ciprian M},
  journal={Journal of Computational and Graphical Statistics},
  volume={23},
  number={1},
  pages={46--64},
  year={2014},
  publisher={Taylor \& Francis}
}

@article{choi2018bayesian,
  title={Bayesian spatially dependent variable selection for small area health modeling},
  author={Choi, Jungsoon and Lawson, Andrew B},
  journal={Statistical methods in medical research},
  volume={27},
  number={1},
  pages={234--249},
  year={2018},
  publisher={SAGE Publications Sage UK: London, England}
}

@article{abatzoglou2016impact,
  title={Impact of anthropogenic climate change on wildfire across western {US} forests},
  author={Abatzoglou, John T and Williams, A Park},
  journal={Proceedings of the National Academy of Sciences},
  volume={113},
  number={42},
  pages={11770--11775},
  year={2016},
  publisher={National Academy of Sciences}
}

@article{bai2019fast,
  title={Fast algorithms and theory for high-dimensional {B}ayesian varying coefficient models},
  author={Bai, Ray and Boland, Mary R and Chen, Yong},
  journal={arXiv preprint arXiv:1907.06477},
  year={2019}
}

@article{barbieri2004optimal,
  title={Optimal predictive model selection},
  author={Barbieri, Maria Maddalena and Berger, James O},
  journal={The Annals of Statistics},
  volume={32},
  number={3},
  pages={870--897},
  year={2004},
  publisher={Institute of Mathematical Statistics}
}

@article{brunsdon1996geographically,
  title={Geographically weighted regression: A method for exploring spatial nonstationarity},
  author={Brunsdon, Chris and Fotheringham, A Stewart and Charlton, Martin E},
  journal={Geographical Analysis},
  volume={28},
  number={4},
  pages={281--298},
  year={1996},
  publisher={Wiley Online Library}
}

@article{comber2024multiscale,
  title={Multiscale spatially varying coefficient modelling using a Geographical {G}aussian Process {GAM}},
  author={Comber, Alexis and Harris, Paul and Brunsdon, Chris},
  journal={International Journal of Geographical Information Science},
  volume={38},
  number={1},
  pages={27--47},
  year={2024},
  publisher={Taylor \& Francis}
}

@article{diffenbaugh2015anthropogenic,
  title={Anthropogenic warming has increased drought risk in {C}alifornia},
  author={Diffenbaugh, Noah S and Swain, Daniel L and Touma, Danielle},
  journal={Proceedings of the National Academy of Sciences},
  volume={112},
  number={13},
  pages={3931--3936},
  year={2015},
  publisher={National Academy of Sciences}
}

@article{finley2011comparing,
  title={Comparing spatially-varying coefficients models for analysis of ecological data with non-stationary and anisotropic residual dependence},
  author={Finley, Andrew O},
  journal={Methods in Ecology and Evolution},
  volume={2},
  number={2},
  pages={143--154},
  year={2011},
  publisher={Wiley Online Library}
}

@book{fotheringham2002geographically,
  title={Geographically Weighted Regression: The Analysis of Spatially Varying Relationships},
  author={Fotheringham, A.S. and Brunsdon, C. and Charlton, M.},
  isbn={9780471496168},
  lccn={2003272388},
  url={https://books.google.com/books?id=GFjFX-3dH68C},
  year={2002},
  publisher={Wiley},
  address={Chichester, England}
}

@article{fotheringham2017multiscale,
  title={Multiscale geographically weighted regression (MGWR)},
  author={Fotheringham, A Stewart and Yang, Wenbai and Kang, Wei},
  journal={Annals of the American Association of Geographers},
  volume={107},
  number={6},
  pages={1247--1265},
  year={2017},
  publisher={Taylor \& Francis}
}

@article{gelfand2003spatial,
  title={Spatial modeling with spatially varying coefficient processes},
  author={Gelfand, Alan E and Kim, Hyon-Jung and Sirmans, CF and Banerjee, Sudipto},
  journal={Journal of the American Statistical Association},
  volume={98},
  number={462},
  pages={387--396},
  year={2003},
  publisher={Taylor \& Francis}
}

@article{gelman1992inference,
  title={Inference from iterative simulation using multiple sequences},
  author={Gelman, Andrew and Rubin, Donald B},
  journal={Statistical Science},
  volume={7},
  number={4},
  pages={457--472},
  year={1992},
  publisher={Institute of Mathematical Statistics}
}

@article{george1993variable,
  title={Variable selection via Gibbs sampling},
  author={George, Edward I and McCulloch, Robert E},
  journal={Journal of the American Statistical Association},
  volume={88},
  number={423},
  pages={881--889},
  year={1993},
  publisher={Taylor \& Francis}
}

@misc{guhaniyogi2025bayesiandatasketchingvarying,
  title={Bayesian Data Sketching for Varying Coefficient Regression Models}, 
  author={Guhaniyogi, Rajarshi and Baracaldo, Laura and Banerjee, Sudipto},
  year={2025},
  eprint={2506.00270},
  archivePrefix={arXiv},
  primaryClass={stat.ML},
  url={https://arxiv.org/abs/2506.00270}
}

@article{mukhtar2025elevation,
  title={Elevation-dependent vegetation greening and its responses to climate changes in the south slope of the Himalayas},
  author={Mukhtar, Hamza and Yang, Yujia and Xu, Mengjiao and Wu, Jiujiang and Abbas, Sawaid and Wei, Da and Zhao, Wei},
  journal={Geophysical Research Letters},
  volume={52},
  number={4},
  pages={e2024GL113276},
  year={2025},
  publisher={Wiley Online Library}
}

@article{pettorelli2005using,
  title={Using the satellite-derived NDVI to assess ecological responses to environmental change},
  author={Pettorelli, Nathalie and Vik, Jon Olav and Mysterud, Atle and Gaillard, Jean-Michel and Tucker, Compton J and Stenseth, Nils Chr},
  journal={Trends in Ecology \& Evolution},
  volume={20},
  number={9},
  pages={503--510},
  year={2005},
  publisher={Elsevier}
}

@article{pickett1995landscape,
  title={Landscape ecology: Spatial heterogeneity in ecological systems},
  author={Pickett, Steward TA and Cadenasso, Mary L},
  journal={Science},
  volume={269},
  number={5222},
  pages={331--334},
  year={1995},
  publisher={American Association for the Advancement of Science}
}

@article{running2004continuous,
  title={A continuous satellite-derived measure of global terrestrial primary production},
  author={Running, Steven W and Nemani, Ramakrishna R and Heinsch, Faith Ann and Zhao, Maosheng and Reeves, Matt and Hashimoto, Hirofumi},
  journal={Bioscience},
  volume={54},
  number={6},
  pages={547--560},
  year={2004},
  publisher={American Institute of Biological Sciences}
}

@article{sun2024alternating,
  title={Alternating dominant effects of temperature and precipitation along elevational gradient on the alpine and subalpine vegetation activities in southwestern China},
  author={Sun, Meirong and Sun, Pengsen and Liu, Ning and Zhang, Lei and Yu, Zhen and Feng, Qiuhong and Smettem, Keith and Liu, Shirong},
  journal={Forest Ecology and Management},
  volume={554},
  pages={121668},
  year={2024},
  publisher={Elsevier}
}

@article{turner2005landscape,
  title={Landscape ecology: What is the state of the science?},
  author={Turner, Monica G},
  journal={Annual Review of Ecology, Evolution, and Systematics},
  volume={36},
  number={1},
  pages={319--344},
  year={2005},
  publisher={Annual Reviews}
}

@article{wheeler2005multicollinearity,
  title={Multicollinearity and correlation among local regression coefficients in geographically weighted regression},
  author={Wheeler, David and Tiefelsdorf, Michael},
  journal={Journal of Geographical Systems},
  volume={7},
  number={2},
  pages={161--187},
  year={2005},
  publisher={Springer}
}

@article{williams2019observed,
  title={Observed impacts of anthropogenic climate change on wildfire in California},
  author={Williams, A Park and Abatzoglou, John T and Gershunov, Alexander and Guzman-Morales, Janin and Bishop, Daniel A and Balch, Jennifer K and Lettenmaier, Dennis P},
  journal={Earth's Future},
  volume={7},
  number={8},
  pages={892--910},
  year={2019},
  publisher={Wiley Online Library}
}

@article{williams2020large,
  title={Large contribution from anthropogenic warming to an emerging North American megadrought},
  author={Williams, A Park and Cook, Edward R and Smerdon, Jason E and Cook, Benjamin I and Abatzoglou, John T and Bolles, Kasey and Baek, Seung H and Badger, Andrew M and Livneh, Ben},
  journal={Science},
  volume={368},
  number={6488},
  pages={314--318},
  year={2020},
  publisher={American Association for the Advancement of Science}
}

@article{wood2016smoothing,
  title={Smoothing parameter and model selection for general smooth models},
  author={Wood, Simon N and Pya, Natalya and S{\"a}fken, Benjamin},
  journal={Journal of the American Statistical Association},
  volume={111},
  number={516},
  pages={1548--1563},
  year={2016},
  publisher={Taylor \& Francis}
}

@article{xu2015bayesian,
  title={Bayesian variable selection and estimation for group lasso},
  author={Xu, Xiaofan and Ghosh, Malay},
  journal={Bayesian Analysis},
  volume={10},
  number={4},
  pages={909--936},
  year={2015},
  publisher={International Society for Bayesian Analysis}
}

@article{yuan2006model,
  title={Model selection and estimation in regression with grouped variables},
  author={Yuan, Ming and Lin, Yi},
  journal={Journal of the Royal Statistical Society Series B: Statistical Methodology},
  volume={68},
  number={1},
  pages={49--67},
  year={2006},
  publisher={Oxford University Press}
}

\end{document}